\documentclass[a4paper,twocolumn,english,superscriptaddress,showpacs,showkeys]{revtex4}
\usepackage{times}
\usepackage[T1]{fontenc}
\usepackage[latin1]{inputenc}
\usepackage{graphicx}
\usepackage{amssymb}

\makeatletter

\usepackage{babel}
\makeatother
\begin{document}

\title{Electron diffraction analysis of individual single-walled carbon
nanotubes}

\author{Jannik C. Meyer}

\email{j.meyer@fkf.mpg.de}

\affiliation{Max Planck Institute for Solid State Research, Stuttgart, Germany}

\author{Matthieu Paillet}

\affiliation{Laboratiore des Colloides, Verres et Nanomateriaux, Universite de
Montpellier II, France}

\author{Georg S. Duesberg}

\affiliation{Infineon Technologies Corporate Research, Munich, Germany}

\author{Siegmar Roth}

\affiliation{Max Planck Institute for Solid State Research, Stuttgart, Germany}

\begin{abstract}
We present a detailed electron diffraction study of individual single-walled
carbon nanotubes. A novel sample preparation procedure provides well-separated,
long and straight individual single-shell nanotubes. Diffraction experiments
are carried out at 60kV, below the threshold for knock-on damage in
carbon nanotubes. We describe experimental parameters that allow single-tube
electron diffraction experiments with widely available thermal emission
transmission electron microscopes. Further, we review the simulation
of diffraction patterns for these objects.
\end{abstract}



\maketitle

\section{Introduction}

From a diffraction analysis, it is possible to derive the exact lattice
structure of an individual single-walled carbon nanotube (SWNT). The
nanotube sections used for electron diffraction experiments in a transmission
electron microscope (TEM) contain only a small number of identical
atoms that interact weakly with the electron beam. Therefore the diffraction
patterns can be easily understood and simulated by simple approximations.
For the same reasons, obtaining a diffraction pattern is an experimental
challenge. Both the simulation of diffraction patterns, and diffraction
experiments at acceleration voltages below the knock-on damage threshold
are described. The latter is important for non-destructive determination
of the nanotube structure, required for electron diffraction experiments
in combination with other single-tube measurements on the same nanotube.

Detailed diffraction studies exist on multi-walled \cite{XFZhangMWED93,VanLanduytAmelinckxReciprSpOfMWNTs94,LucasCNTelDiffr97,AmelinckxED98}
and double-walled \cite{KociakDWNTaccuracy03,LinkingNMandTransportDWNTwPiezo,ZuoDWNTOversampling03,BondStretchingInDWNT}
carbon nanotubes, and bundles of single-walled carbon nanotubes \cite{CowleySTEM_ED97,QinSWNTBundleED97,CowleyED97,LambinZigzagBundle97,BernaertsED97,ChenED98,ZhuED99,HenrardSW_ED00,ColomerSWNTUniqueHel01,ColomerSW_ED02,ColomerSWNTUniqueHel04}.
Diffraction experiments on \emph{individual} SWNTs are an experimental
challenge because they require finding sufficiently long, straight
sections of an individual nanotubes (raw material is mostly bundled)
that are stable throughout the exposure. Nanotubes with more than
one shell are usually stiff enough to exhibit long, straight sections
with standard sample preparation procedures. SWNTs, on the other hand,
tend to curve, and to combine into bundles. A curved nanotube is not
a periodic one-dimensional structure. Few examples of sharp diffraction
patterns from individual SWNTs exist in the literature \cite{IijimaSWNT1993,IijimaHelicity96,GaoSWNTnanoareaED03}.
However with the sample preparation procedure shown in this work,
long straight sections of individual SWNTs are easily obtained. We
have developed a way to suspend nanotubes in a metal grid that provides
long and straight tube sections. In the resulting samples, tubes suitable
for diffraction are easily found and reliably produce a diffraction
pattern with a close to normal incidence. With the given conditions,
more than 50\% of the diffraction exposures result in a pattern that
can be uniquely assigned to a nanotube structure. The {}``failed''
exposures are attributed to curved or strongly vibrating tubes, or
to objects other than the tube illuminated by the beam. Indices for
more than 50 nanotubes were determined so far. It is a reliable procedure,
which is a prerequisite for the combination with other experiments
on the same tube. We have shown the combination of TEM imaging and
transport measurements previously \cite{MeyerTransportTEM04}, and
plan to extend this method towards transport and electron diffraction
on the same molecule. 

For the simulated diffraction images, a computer program was written
which produces the nanotube structure for given indices, and then
calculates the diffraction intensities using the principles and equations
shown in the simulations section. We determine the nanotube structure
by comparison with simulated images, and find that there is exactly
one nanotube structure which matches the experimental pattern. For
an analytical analysis of the diffraction patterns based on the helical
structure, see e.g. \cite{LambinHelixCalc96,LambinZigzagBundle97,LucasCNTelDiffr97,ElDiffTheory94,QinEDtheory98}.
We note that the index determination solely from relative peak distances
as described in \cite{QinSWNTfromLayerLineCPL05,Qin4WNTapl05} is
valid only for precisely normal incidence, a condition that is not
easily established experimentally. The comparison with simulations,
in contrast, allows to determine the indices and at the same time
to measure the incidence angle by varying these parameters until the
simulated pattern matches the experimental one.

Other techniques for identifying the structural indices are atomic-resolution
TEM imaging \cite{IijimaAtomicDefects04,AtomicCorrDWNT05,GreenPhaseReconstr03},
Raman spectroscopy \cite{TelgPRL04,JorioPRL01}, fluorescence spectroscopy
\cite{BachiloPLmapSci02,BachiloJACS03,MaruyamaCPL04}, and scanning
tunnelling microscopy and spectroscopy (STM/STS) \cite{DekkerSTMnat98,LieberSTMnat98,DekkerSTMPRB00,VitaliSTSPRL04}.
Electron diffraction and high-resolution images can be directly related
to a pair of carbon nanotube structural indices (n,m), while spectroscopic
techniques rely on a modelization of electronic and vibrational properties
of carbon nanotubes. A quantitiative analysis of the index distribution
is complicated by a different response for different nanotubes both
in Raman \cite{MachonPRB05} and fluorescence spectroscopy. Further,
fluorescence spectroscopy detects only semiconducting nanotubes.

\section{Sample preparation}

Carbon nanotubes are grown by chemical vapour deposition (CVD) on
highly doped Si substrates with a 200nm oxide layer. Different CVD
growth conditions were used, as described in \cite{PailletCVDJPCB04,PailletDRMCVD05,DuesbCVD04}.
A grid structure is prepared on top of the nanotubes by electron beam
lithography, thermal evaporation of 3nm Cr and 110nm Au, and a lift-off
process. The sample is cleaved so that the structure is on the cleaved
edge of the substrate (Fig. \ref{cap:Sample-preparation}a). There
are two possible etching processes to obtain the free-standing structure
(Fig. \ref{cap:Sample-preparation}b): 

\begin{figure}[!h!t]
\includegraphics[%
  width=1.0\linewidth]{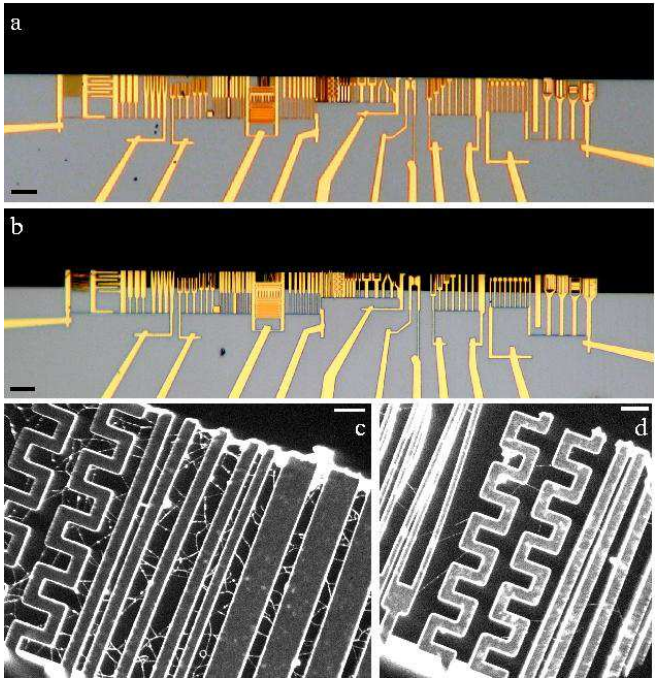}

\caption{Sample preparation. (a) The grid structure, prepared by electron
beam lithography on top of the nanotube network, close to the cleaved
edge of the substrate. (b) After etching, part of the grid is free-standing.
This free-standing part can be observed by TEM. (c) shows a dark-field
mode TEM image of a high-density sample, (d) a low-density sample.
Especially in lower-density samples, long straight individual SWNTs
of high purity (depending on CVD conditions) are obtained. The dark-field
mode is used here since the clean SWNTs are not seen in bright-field
mode at lower magnifications. Scale bars are 10\ensuremath{µ}m in
(a) and (b), and 1\ensuremath{µ}m in (c) and (d). \label{cap:Sample-preparation}}
\end{figure}

\begin{enumerate}
\item The more simple process is a 6 hours etch in 30\%KOH at 60°C. The
KOH removes the bulk Si, and slowly the oxide layer. Thus, the structure
is undercut mostly from the side. The etching process has to be stopped
when the oxide layer is completely removed in the free-standing part,
but still present on the substrate. The etch rate of the bulk Si can
be controlled by electrically biasing the substrate with respect to
the etching solution. The etch process is monitored with an optical
microscope, and the bias is switched to a positive potential on the
bulk Si as soon as the free-standing part has the desired width. This
stops the etching on the bulk Si, while the etching on the oxide layer
continues.
\item A two-step etch process consists of a TMAH etch (15\%, 60°C) to underetch
the structure and oxide layer from the side of the cleaved edge, followed
by a buffered HF etch (6.5\%, 2min) to remove the oxide layer. The
width of the free-standing part can be controlled by the time of the
TMAH etch (typically 1-3 hours). This process allows the preparation
of more {}``fragile'' structures, like large free-standing patterns,
very long suspended nanotubes, or lithographically defined metal objects
suspended on carbon nanotubes. The TMAH etch step is monitored with
an optical microscope, and stopped when the free-standing part has
reached a sufficient width.
\end{enumerate}
Both processes are followed by transferring the sample into water,
then into isopropanol, and finally into acetone (VLSI grade). A specially
designed sample holder prevents the edge of the sample from drying.
Finally, a critical point drying step with carbon dioxide is carried
out. The substrates are small enough to be glued into conventional
TEM rings. The etching process does not damage the nanotubes: The
diffraction patterns show a well preserved crystallinity, and the
Raman spectra \cite{MeyerEDRam05} show no unusual features and a
low D line.

\begin{figure}[!h!t]
\includegraphics[%
  width=1.0\linewidth]{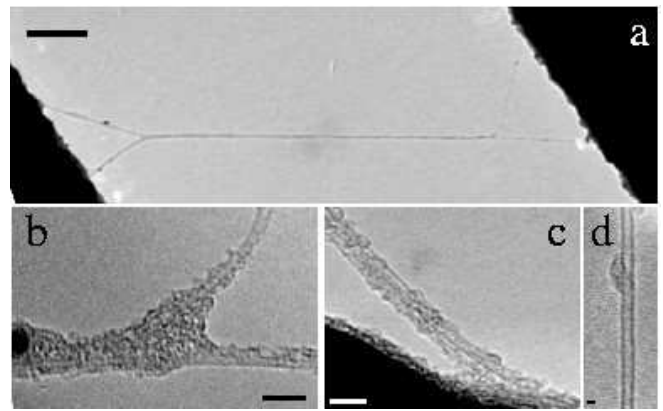}

\caption{(a) Stretching of a small bundle by coalescence of the two nanotubes.
We often observe small bundles splitting near the contact structure.
These nanotubes were most likely separated before the etching process,
but close together. They are fixed at the contacts, but the central
part is free to move after etching. Due to van der Waals attraction
the central sections have formed a bundle. As a result, the tube sections
are straightened. (b) is a close-up on the right splitting point in
(a). (c) shows a nanotube end (a bundle in this case) close to the
contact. Tubes often appear curved just at the ends, suggesting that
they are stretched by an attractive force towards the contacts. Most
of the amorphous carbon is deposited during the TEM analysis: (d)
shows a nanotube section after quickly zooming in to a high magnification.
Scale bars are 100nm (a), 5nm (b+c) and 1nm (d). \label{cap:Stretching} }
\end{figure}

As a result of this procedure, we obtain well-defined nanotube samples
with nanotubes nearly orthogonal to the beam. In low-density samples
(Fig. \ref{cap:Sample-preparation}d), approximately 50\% of the nanotubes
are individual and well stretched between the contacts for a diffraction
analysis. In higher-density samples (Fig. \ref{cap:Sample-preparation}c),
bundling occurs more frequently, and curved free-standing bundles
appear. Individual tubes appear to be stretched by van der Waals attraction
towards the contacts. Especially if they are not ending orthogonal
to the contact, they curve towards a parallel alignment with the contact
edge. Similarly, stretching is observed in small bundles (which are
most likely formed after etching, when the tubes are free to move).
An example is shown in Fig. \ref{cap:Stretching}.

We want to point out that carbon nanotubes, grown by CVD directly
on a bulk substrate, can not usually be investigated by TEM and electron
diffraction. They become accessible only due to our sample preparation
procedure. This permits a quality control of CVD grown nanotubes,
and provides information not available from AFM or SEM investigations
about bundling, precise diameters, number of shells, and amorphous
carbon coating from TEM, and index distribution from diffraction analysis.

Instead of CVD grown nanotubes, it is also possible to use nanotubes
deposited on the substrates from a suspension. However we find that
this leads to a high amount of SWNT bundles.

\section{Simulations}

Two different ways to calculate the diffraction pattern of a SWNT
are used. The real-space path summation approach is computationally
expensive but easy to understand and implement. It produces the correct
peak positions, but not the right intensities. Our determination of
the nanotube structure only depends on the peak positions. The alternative
is computing the Fourier transform of the projected atomic potentials.
This is the standard approach for thin objects that do not require
multi-slice algorithms. However, peak positions deviate slightly from
the correct values because the curvature of the Ewald sphere is not
taken into account.

Individual single-walled carbon nanotubes are one of the few systems
of which the interaction with electrons in a TEM is well described
by the weak phase object approximation (WPOA) and the 1st Born approximation.
We use non-relativistic quantum mechanics with relativistically corrected
values for the electron mass $m$ and wavelength $k$. It has been
shown that electron microscopic problems are well described in this
way \cite{relTEM62,relTEM79,Spence_HRTEM}.

\subsection{Path summation approach\label{sub:Path-summation-approach}}

\begin{figure}[!h!t]
\begin{center}\includegraphics[%
  width=0.65\linewidth]{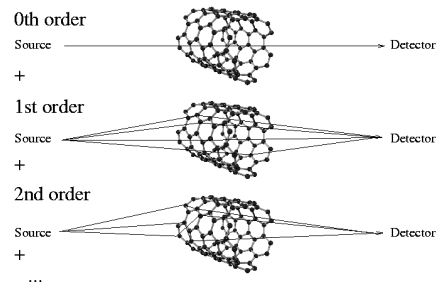}\end{center}

\caption{Contributions to the path summation, sorted by the number of interactions.
The phases $e^{ikr}$along each path are summed up. It is equivalent
to the 1st Born approximation to consider only 0th and 1st order contributions.
The summation is carried out for every point in the simulated image.\label{cap:BornContrib}}
\end{figure}

Due to the small number of atoms in our one-dimensional samples it
is possible to numerically calculate diffraction images from basic
physical principles in a real-space representation. We consider an
electron propagating from $(r_{0},t_{0})$ to $(r_{1},t_{1})$. In
the most general case, the complex probability amplitude $\Psi(r_{1},t_{1})$
is given by a summation over all paths starting from $\Psi(r_{0},t_{0})$
\cite{FeynmanPathint1948,FeynmanQMbook1965}:

\begin{equation}
\Psi(r_{1},t_{1})=\Psi(r_{0},t_{0})\cdot\sum_{\textrm{paths}}e^{\frac{i}{\hbar}S}\label{eq:summationGeneral}\end{equation}

where $S$ is the \emph{action} along the path. For propagation in
free space, the summation can be replaced the term for the classical
path, multiplied with a constant $c_{1}'$ or $c_{1}$ \cite{FeynmanQMbook1965}:

\begin{equation}
\sum_{\textrm{paths}}e^{\frac{i}{\hbar}S}=c_{1}'\cdot e^{\frac{i}{\hbar}S_{cl.}}=c_{1}\cdot\frac{1}{\left|r\right|}e^{i(kr-\omega t)}\label{eq:SummationFreespc}\end{equation}
If we idealize the atoms in an object as point scatterers, we can
divide the summation into the scattering events and piecewise propagation
in free space. This is illustrated in Figure \ref{cap:BornContrib}.
We consider the propagation from the source to the detector as the
sum of the direct contribution, plus the paths including a single
scattering event, two events and so on. The sum is thus written as:

\begin{eqnarray}
\sum_{\textrm{paths}}e^{\frac{i}{\hbar}S} & = & c_{1}\cdot\frac{1}{\left|r_{SD}\right|}e^{ikr_{SD}}\nonumber \\
 & + & c_{1}^{2}\cdot\sum_{n}\frac{1}{\left|r_{SA_{n}}\right|\left|r_{A_{n}D}\right|}e^{ikr_{SA_{n}}}f_{n}e^{ikr_{A_{n}D}}\nonumber \\
 & + & c_{1}^{3}\cdot\sum_{n}\sum_{m}\,...\label{eq:BornSeries1}\end{eqnarray}

Here, $r_{SD}$ is the distance from the source to the detector, $r_{SA_{n}}$
the distance from the source to atom $n$, and $r_{A_{n}D}$ the distance
from atom $n$ to the detector. The summations are over all atoms
of the object, $f_{n}$ is the factor describing the scattering event
at atom $n$. For a set of identical atoms, the actual value of the
scattering factor $f_{n}$ is not important. The above summation represents
Huygens principle of a spherical wave originating from every atom
in the sample. Considering only the 0th and 1st order contributions,
i.e. lines 1 and 2 in the above summation, is the equivalent of the
1st Born approximation for atoms idealized as point scatterers.

For objects containing a small number of atoms, the above summation
can be carried out numerically. Given the small number of light (i.e.,
weakly scattering) atoms in an individual single-walled carbon nanotube,
the 1st order approximation is reasonable. Assuming a detector at
a large distance from the nanotube, diffraction patterns can be calculated
numerically directly from (\ref{eq:BornSeries1}). Although reciprocal-space
approaches for image simulation (shown below) are more efficient in
terms of required computing time, the real-space summation can conveniently
be carried out on a standard PC for the small number of atoms in a
carbon nanotube. The length of the simulated nanotube section determines
the width of the peaks. A 50nm long simulated nanotube is sufficient
for calculating a high-quality diffraction image.

\subsection{Fourier space approach\label{sub:Fourier-space-approach}}

Alternatively, diffraction patterns of carbon nanotubes can be calculated
from single-slice projected potentials. This is a widely used method
for sufficiently thin objects that do not require multi-slice algorithms.
Detailed descriptions are found in \cite{Spence_HRTEM,KirklandEMcomputing,Buseck_HRTEMandAssoc,HumphreysElScat79,Peng99elasticScat}.
Here we shortly summarize the approximations and calculations we use. 

Within the 1st Born approximation the diffraction pattern can be obtained
from a Fourier transform of the scattering potential $V(r)$. In the
case of small momentum transfer, we can approximate the observed section
of the Ewald sphere by a planar section. In this approximation the
diffraction pattern is the 2D-Fourier transform of the projected scattering
potential. 

\begin{figure}[!h!t]
\includegraphics[%
  width=1.0\linewidth]{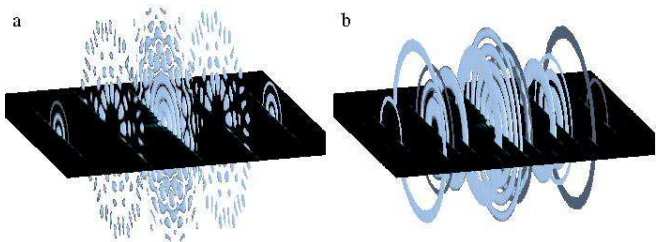}

\caption{Three-dimensional Fourier transformation of a (7,7) nanotube (a)
and an (8,3) nanotube (b). An isosurface is drawn around the higher
intensity volume elements, shown in a perspective view. For a bulk
crystal, this image would show the reciprocal lattice points. But
since we have a one-dimensional structure, we obtain features that
are extended in two dimensions: The diffraction intensities are cumulated
on discs orthogonal to the tube axis. The planar section corresponds
to the section of the Ewald sphere that would be observed in an electron
diffraction experiment with normal incidence on the tube. For the
(7,7) nanotube, a discrete 14-fold rotational symmetry around the
tube axis is present in the Fourier transform. \label{cap:FT3D}}
\end{figure}

In Figure \ref{cap:FT3D} the three-dimensional Fourier transform
of a carbon nanotube is calculated, and a planar cut corresponding
to an observable section of the Ewald sphere is shown. Since the nanotube
is a periodic one-dimensional structure, the intensity in reciprocal
space is cumulated on discs orthogonal to the tube axis. The spacing
of these discs corresponds to the periodicities along the tube axis,
while the radial intensity distribution is determined by the tube
diameter. The armchair and zigzag nanotubes exhibit a discrete rotational
symmetry, which is also present in the Fourier transform. Therefore,
the diffraction pattern would change if the tube is rotated around
its axis.

The scattering potential $V(r)$ is calculated from the tabularized
values of Gaussian fits to relativistic Hartree-Fock calculations
given in \cite{DoyleTurner68}. This derivation can be found e.g.
in \cite{Peng99elasticScat}. Within the WPOA (and planar approximation
of the Ewald sphere), the diffraction pattern is calculated from a
Fourier transform of the projected potential. Alternatively, the phase
shift in the exit plane wave function can be calculated from the projected
potential (phase object approximation). The diffraction pattern is
then calculated from a Fourier transform of the exit plane wave function. 

Computing diffraction patterns by a fast Fourier transformation of
the exit plane wave function is much faster than the path summation
calculation shown in section \ref{sub:Path-summation-approach}. The
fast Fourier transform typically takes less than a minute, while the
real space summation needs several hours. However, it implies approximating
a section of the Ewald sphere as a plane.

\subsection{Qualitative description of the diffraction pattern}

The diffraction pattern of a single-walled carbon nanotube can be
well separated into features that depend on the nanotube diameter,
and others which depend on the rolling angle of the graphene sheet.
The most prominent feature is the equatorial line, which is similar
to a double-slit interference pattern. The periodicity of the intensities
on this line is related only to the nanotube diameter. 

\begin{figure}[!h!t]
\includegraphics[%
  width=1.0\linewidth]{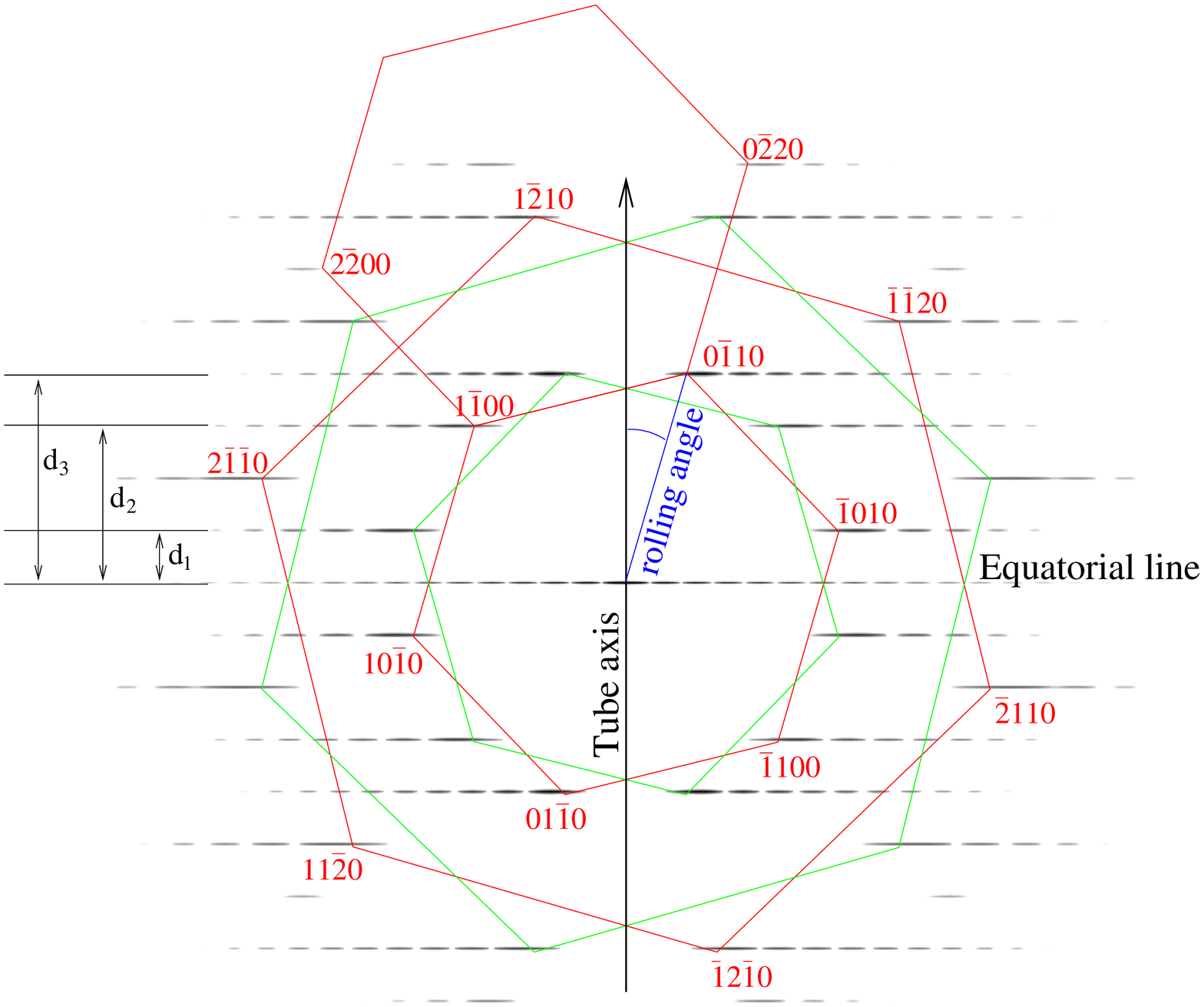}

\caption{Orientation of the reciprocal graphene lattices in the SWNT diffraction
pattern. Two sets of peaks that correspond to the top and bottom graphene
layer in the nanotube are visible. The Miller-Bravais indices for
one of the two hexagonal lattices are given. A rolling angle of 0°
corresponds to a zig-zag nanotube, 30° represents an armchair tube.
It is not recommended to measure the angle as indicated in the diagram,
since the indicated angle is only true for an exactly normal incidence
of the beam on the tube. Instead, the angle can be reliably calculated
from the relative distances of the peaks to the equatorial line, indicated
by $d_{1}$, $d_{2}$ and $d_{3}$. The streaked shape of the diffraction
peaks is due to the curvature of the graphene sheet. \label{cap:GrapheneInED}}
\end{figure}

From further peaks in the pattern we can determine the orientation
of the (reciprocal) graphene hexagonal lattices. A simulated diffraction
pattern for a (15,06) nanotube, and a part of the reciprocal graphene
lattice, is shown in Figure \ref{cap:GrapheneInED}. Two separate
sets of peaks that correspond to the top and bottom graphene layer
are visible for chiral nanotubes (which coincide for armchair and
zigzag nanotubes). The peaks appear as streaks due to the curvature
of the graphene sheet.

The orientation of the graphene lattice, i.e. the rolling angle of
the nanotube, can be calculated independent of the incidence angle
from the relative distances of the peaks from the equatorial line.
This measure is also independent of the scale (camera length) or a
diffraction astigmatism. Thus, the rolling angle can be reliably and
precisely measured with a precision of up to $0.1°$. Using e.g. the
distances $d_{1}-d_{3}$ in Figure \ref{cap:GrapheneInED}, the rolling
vector is \cite{GaoSWNTnanoareaED03}:

\begin{equation}
\alpha=\arctan\left(\frac{1}{\sqrt{3}}\cdot\frac{d_{2}-d_{1}}{d_{3}}\right)=\arctan\left(\frac{1}{\sqrt{3}}\cdot\frac{2d_{2}-d_{3}}{d3}\right)\label{eq:AngleFromDist}\end{equation}

For identifying the indices of a nanotube from a diffraction pattern,
a measurement of the diameter and angle from the pattern as described
above can be used as a starting point. Then, the simulated images
are needed. Once a matching structure is found, the most important
aspect for an unambiguous identification is to \emph{exclude} all
other possible nanotube structures (within a reasonable range of angle
and diameter). For a sufficiently sharp diffraction pattern, there
is usually only one set of indices (n,m) for which the simulated diffraction
pattern matches the experimental data. Only for very large diameter
nanotubes, the parameters (diameter and angle) of the different candidates
are spaced so close together that they can not be distinguished.

\subsection{Accuracy of the simulation methods }

The single-walled carbon nanotube is one of the few systems well described
by the weak phase object approximation (WPOA). In addition to calculations
\cite{LucasCNTelDiffr97}, the validity of the WPOA is demonstrated
in an impressive way by the oversampling and iterative phase retrieval
for a nanotube from diffraction intensities \cite{ZuoDWNTOversampling03}.
The reconstruction in \cite{ZuoDWNTOversampling03} is based on the
WPOA, and has worked for the large double-walled carbon nanotube.
However the iterative phase retrieval is not possible from our diffraction
data, because it requires the fully coherent illumination of a short
nanotube section which is available only with field-emission electron
sources. 

From a Fourier transformation of the projected potential, one can
quickly calculate a simulated diffraction pattern (e.g. Figure \ref{cap:GrapheneInED}).
This approach approximates the visible section of the Ewald sphere
as a plane. The computationally more expensive path summation approach
(section \ref{sub:Path-summation-approach}) naturally includes the
curvature. The curvature effect is more pronounced at lower acceleration
voltages. The difference can be neglected for normal incidence, but
it is clearly seen in diffraction images for non-normal incidence.
For non-normal incidence, the peak positions are no longer symmetric
with respect to the origin. The actual difference between curved and
planar approximation to the Ewald sphere depends on the tube structure
and actual incidence angle. We find that for an incidence angle within
$\approx$10° of normal incidence, a safe assignment of nanotube indices
using the planar approximation to the Ewald sphere is possible.

\section{Diffraction images}

\subsection{Experimental procedure}

All SWNT diffraction patterns were obtained using a Zeiss 912$\Omega$
transmission electron microscope. It is equipped with a (thermal)
$\textrm{LaB}_{6}$ electron gun, a Köhler illumination system, and
an energy filter. Images can be taken on two CCD cameras with different
fields of views. The diffraction patterns are recorded on image plates.
The image plates provide a very high sensitivity and dynamic range,
which is not matched by any CCD camera. CCD cameras, and also conventional
film, suffer from a {}``blooming'' effect: Intensity from strongly
excited, saturated pixels spreads out into the nearby regions, occulting
weak intensity diffraction peaks. 

\begin{figure}[!h!t]
\includegraphics[%
  width=1.0\linewidth]{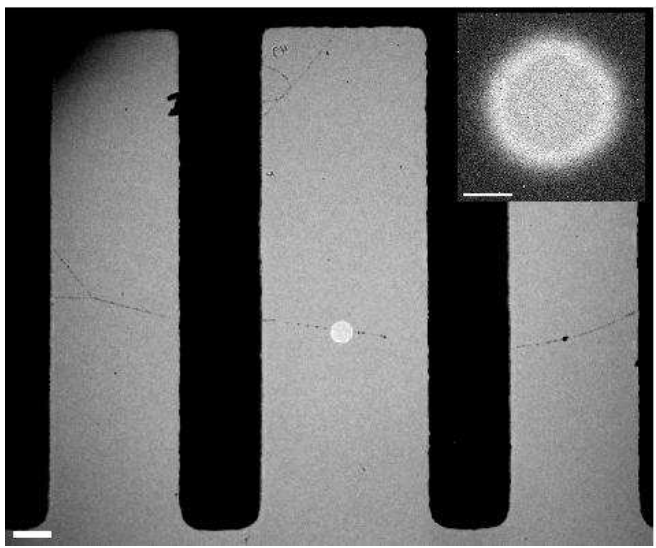}

\caption{Inset: An individual SWNT within the illuminated region of the smallest
condenser aperture. Main image: Overlay of a TEM image of the fully
illuminated sample with an image of smallest condenser aperture (bright
spot in the center). This overlay shows the illuminated section of
the nanotube (for a diffraction pattern) in relation to the sample
structure. The illuminated region for the smallest condenser aperture
has a diameter of 130nm. Scale bar is 200nm, and 50nm for the inset.\label{cap:SWNTinC5ap}}
\end{figure}

Unless noted otherwise, the following procedure and conditions ({}``standard
conditions'') are used for obtaining diffraction patterns. The microscope
is operated at an acceleration voltage of only 60kV, clearly below
the threshold for knock-on damage in carbon nanotubes which is 87kV
\cite{KnockOnDamageJAP01}. The Köhler illumination condition is used
with an illumination angle between 0.1 and 0.2 mrad. The illumination
is limited to a straight section of the carbon nanotube using the
condenser aperture. The smallest condenser aperture, which has a diameter
of 5\ensuremath{µ}m, produces an illuminated region (demagnified image
of the aperture) with a diameter of approximately 130nm. An image
of a carbon nanotube illuminated under these conditions is shown in
Figure \ref{cap:SWNTinC5ap}. Before switching to diffraction mode,
the focus is tuned to the minimum contrast condition. The energy filter
is set to a width of $15-20\textrm{eV}$. The diffraction image is
recorded on image plates with a camera length of 450mm or 580mm and
exposure times of 4 or 5 minutes.

As a reference, diffraction images are recorded under the same conditions,
except for a shorter exposure time, from the metal contacts. These
contacts are polycrystalline, producing Debye-Scherrer ring type of
diffraction patterns. From these, any diffraction astigmatism remaining
after alignment can be detected and compensated, and a precise scale
is obtained. An astigmatism of a few percent may already lead to an
incorrect assignment of the nanotube indices.

\subsection{Discussion of experimental parameters}

Previous diffraction work on individual single-walled carbon nanotubes
\cite{GaoSWNTnanoareaED03} was performed in transmission electron
microscopes equipped with a field-emission electron source, resulting
in highly coherent illumination. Therefore, diffraction images could
be recorded with short exposure times, probably before significant
damage to the nanotubes occurred. However, it can not be excluded
that surface reconstruction and dimensional changes due to the electron
irradiation \cite{AjayanSurfReconstrPRL98} change the nanotube structure
during the exposure.

Using a thermal emitter, we need longer exposure times. Operation
below the threshold for knock-on damage makes it possible to have
stable conditions throughout the long exposure times. It is possible
to obtain several diffraction images from the same nanotube section
without loss of quality. Further, minimizing damage is important if
transport or Raman investigations are to be carried out on the same
nanotube.

\begin{figure}[!h!t]
\begin{center}\includegraphics[%
  height=0.80\textheight]{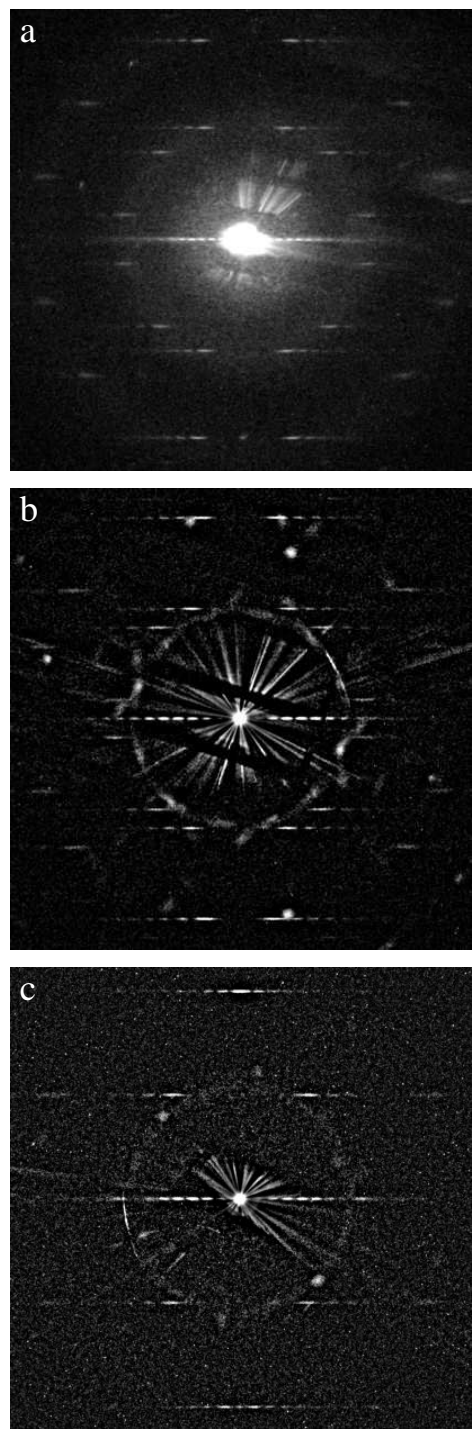}\end{center}

\caption{Examples of diffraction patterns. (a) (24,11) nanotube, (b) (16,09)
nanotube, (c) (13,13) {}``armchair{}`` nanotube. A background subtraction
was done on images (b) and (c). \label{cap:EFandSAeffects}}
\end{figure}

Figure \ref{cap:EFandSAeffects} shows three diffraction patterns
obtained on different nanotubes. All diffraction images obtained in
this way show intense radial lines starting from the central peak.
These are due to electrons scattered into high angles at the condenser
aperture. They are partially shadowed by the structure of the sample.
A shadow of the sample structure is visible in the diffraction pattern
due to this effect.

With a sufficiently sharp diffraction pattern, an unambiguous assignment
of the nanotube structure, or nanotube indices (n,m), is possible.
A straight, clean nanotube is required - a curved nanotube is not
a 1D periodic structure. In the high-resolution images of suspended
nanotubes we are able to obtain a sharp image only from the ends,
due to vibrations of the central part. Such vibrations are not a problem
for diffraction analysis: A \emph{translation} of the nanotube does
not change the diffraction pattern. Only if the vibrations are so
large that the \emph{orientation} of the tube varies, it leads to
a blurred diffraction image.

Although all diffraction images were recorded at the same microscope,
it seems that the key ingredients for a single-tube diffraction analysis
are the small condenser aperture, image plates for recording of the
diffraction patterns, and very straight nanotubes. The condenser aperture
limits the illumination to the region of interest (the nanotube section).
Selecting the area downstream of the sample (using \emph{only} the
selected area aperture) fails, since many other contributions (e.g.
reflections from the metal contact, inelastic contributions) still
reach the detector. The smallest unwanted contributions will occult
the very weak diffraction intensities from the small number of atoms
in our nanotube. We have seen that the energy filter is not necessary,
but improves the image quality. The Köhler illumination condition
provides a homogeneous incident beam; however, convergent-beam electron
diffraction (shown below) is also possible on individual nanotubes.

\subsection{Index assignment}

For the analysis, the diffraction images are rotated to have the equatorial
line in a horizontal direction. If any astigmatism is found in the
reference patterns, it is compensated by rescaling and shearing the
images by the appropriate amount. In some cases a background subtraction
is useful. The above steps can be done with the ImageJ software and
The Gimp (there is a version capable of manipulating 16 and 32-bit
images, called FilmGimp).

\begin{figure}[!h!t]
\includegraphics[%
  width=0.75\linewidth]{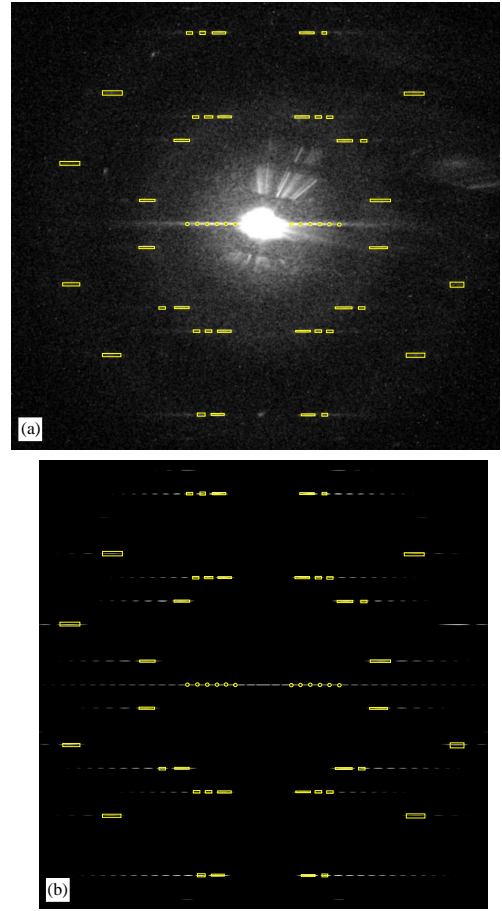}

\caption{Index assignment: (a) The peaks in the experimental diffraction pattern
(from figure \ref{cap:EFandSAeffects}a) are marked. A box is drawn
around the streaked graphene peaks, and in the equatorial line the
minima are marked. In (b), the experimental image is replaced by a
simulated one, without change in the marks. The indices and incidence
angle in the simulation are changed until the pattern matches. Here,
the pattern matches only one pair of indices (n,m)=(24,11). The incidence
angle is within 3° of normal incidence. \label{cap:Index-assignment} }
\end{figure}

The rotated, background subtracted and rescaled images are then imported
in xfig. Here, the features in the image are marked (Figure \ref{cap:Index-assignment}).
Then, simulated images are placed instead of the experimental underneath
the marks, to find those which match. We find that for a sharp diffraction
pattern there is exactly one set of indices (n,m) for which the simulated
image matches the experimental one. The indices and the incidence
angle of the simulated pattern are varied to find the matching set
of parameters. Thus, both the nanotube structure and the incidence
angle are measured from the diffraction pattern. Once a matching pattern
is found, we make sure that all nearby indices do not match, independent
of the incidence angle. By \emph{excluding} all except one pair of
indices (n,m) we obtain an unambiguous assignment. 

The smallest individual carbon nanotube that was identified is a (7,7)
nanotube, which has a diameter of 0.94nm. One of the largest identified
nanotubes is (39,26) which has a diameter of 4.44nm.

\begin{figure}[!h!t]
\includegraphics[%
  width=1.0\linewidth]{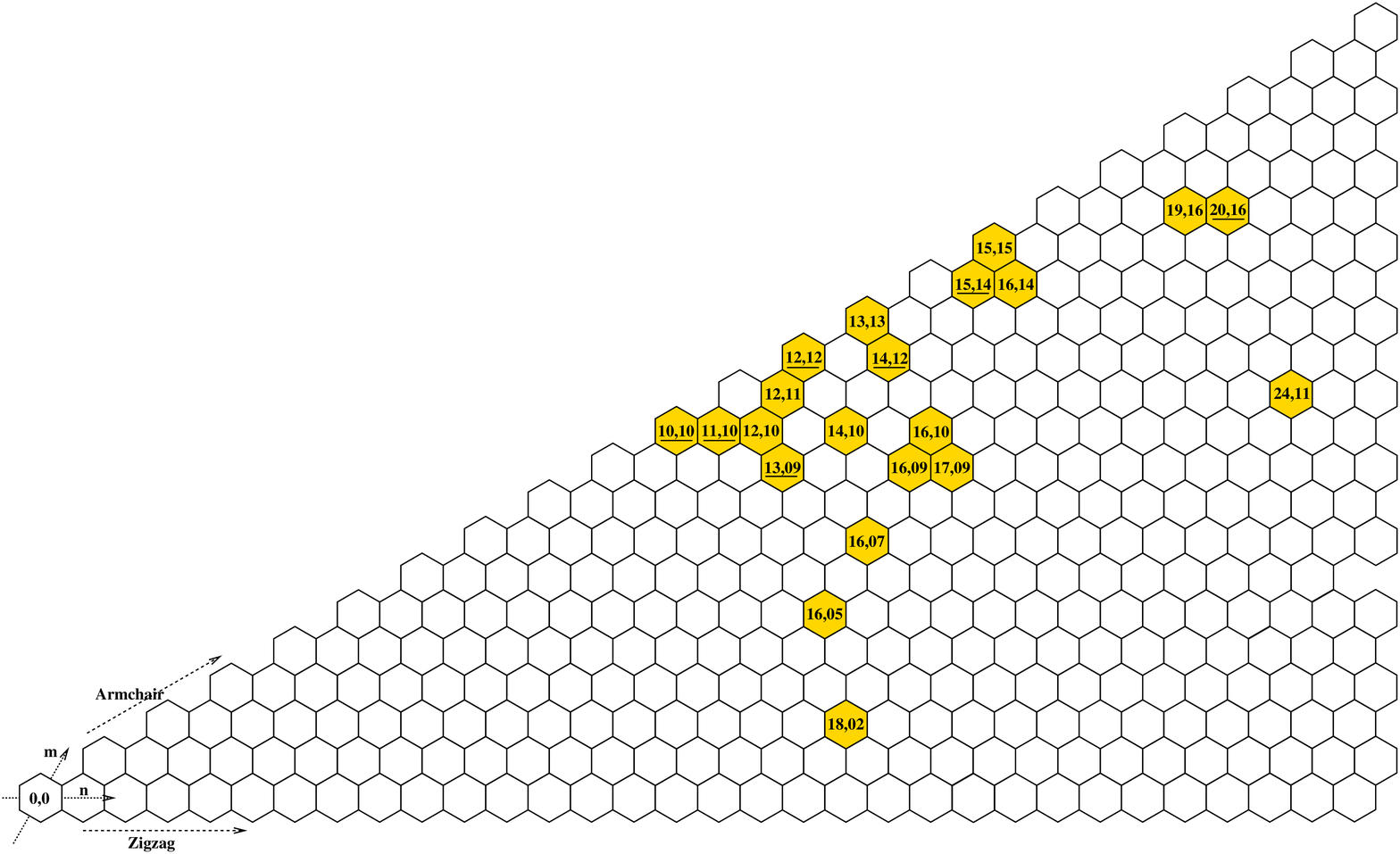}

\caption{Nanotube indices of 28 tubes grown in the same CVD process. The underlined
indices were encountered twice. The rolling angle is not randomly
distributed, but is close to the armchair direction (30°) in the majority
of the nanotubes in this material.\label{cap:CH4stat}}
\end{figure}

Figure \ref{cap:CH4stat} shows the statistics of indices obtained
from nanotubes all grown in the same CVD process. We use 4-5 nm Ni
nanoparticles as catalyst, and methane as carbon feedstock with flows
of 1.2 L/min of methane and 0.6 L/min of H$_{2}$. The synthesis temperature
is 900°C and the duration 10min. We describe the CVD process in detail
in \cite{PailletCVDJPCB04}.

\subsection{Discussion of the index distribution}

Figure \ref{cap:CH4stat} clearly shows that the rolling angle is
not randomly distributed, but is grouped towards the armchair orientation
in this material. This is a direct measurement and it includes metallic
nanotubes. Fluorescence spectroscopic results \cite{BachiloJACS03,MaruyamaCPL04}
show a similar trend, but are limited to semiconducting tubes and
biased by a structure-dependent fluorescence quantum yield. The Raman
study in Ref. \cite{TelgPRL04} finds a more or less random distribution
of nanotube indices in HiPCO nanotubes. Note that all these index
distributions are obtained from nanotubes grown with different methods
and parameters. As demonstrated in \cite{MaruyamaCPL04} the index
distribution strongly depends on growth conditions. We can compare
the angular selectivity in our material with fluorescence studies
of nanotubes produced by alcohol catalytic CVD (ACCVD) under optimized
conditions \cite{MaruyamaCPL04} and of nanotubes grown from solid-supported
Mo/Co catalyst (MoCo) \cite{BachiloJACS03}, under the assumption
of a structure-independent fluorescence yield. An angle of 30° represents
armchair nanotubes, 0° a zigzag tube, and an average of 15° is expected
for a random distribution. The average angle in the MoCo material,
calculated from the fractional intensities listed in \cite{BachiloJACS03},
is 22.8°. The optimized ACCVD appears to yield predominantly (6,5)
(27°) and (7,5) (24.5°), but also significant amounts of (8,3) (15,3°)
and (8,4) (19.1°). The intensities in \cite{MaruyamaCPL04} are given
only in a graphical representation, from which we estimate an average
angle of $\approx$23°. In our sample the mean angle is 25° if we
include all nanotubes, and 24° if we exclude the metallic tubes for
a better comparison with the fluorescence spectroscopic results. Due
to the much larger diameters obtained in our CVD process, there are
many more different nanotube species possible within a given angle
interval. The angular selectivity towards armchair tubes, however,
is at least similar to the samples in \cite{BachiloJACS03} or \cite{MaruyamaCPL04}.

\subsection{Convergent-beam electron diffraction}

Although nanoarea electron diffraction using the Köhler illumination
condition with a small condenser aperture has proven to be very reliable,
we would like to show also the possibility of convergent-beam electron
diffraction (CBED) on individual single-walled carbon nanotubes. Here,
the illuminating electron beam converges to form a small probe of
a few nm on a straight nanotube section. Consequently, the diffraction
pattern consists of extended discs instead of sharp points. It is
possible to investigate even smaller nanotube sections in this way.
An unambiguous nanotube index assignment, however, is not possible
(it could be possible in combination with a very accurate diameter
estimate from a high-resolution image, which is not available in our
case). The image is again recorded on image plates with an exposure
time of five minutes. 

\begin{figure}[!h!t]
\includegraphics[%
  width=1.0\linewidth]{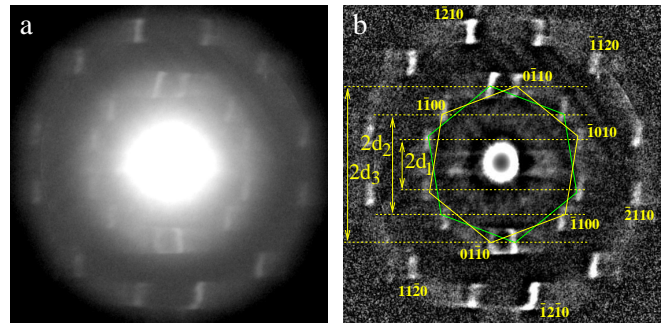}

\caption{Convergent beam electron diffraction (CBED) image of an individual
single-walled carbon nanotube. Original (a) and background subtracted
image (b). The orientation of the graphene lattice is visible, but
the equatorial line is not. The illuminated nanotube section is visible
in each of the diffraction spots. In (b) part of the reciprocal graphene
lattice and indices according to Fig. \ref{cap:GrapheneInED} are
shown. The hexagons do not precisely match the peaks due to non-normal
incidence. The rolling angle of the nanotube can be calculated from
the distances $\textrm{d}_{1}-\textrm{d}_{3}$ according to equation
(\ref{eq:AngleFromDist}), and in this example we obtain 12.5$\pm$0.5°.
\label{cap:CBED}}
\end{figure}

A convergent beam electron diffraction pattern of an individual single
walled nanotube is shown in Figure \ref{cap:CBED}. A distorted image
of the tube is visible in each spot (in fact the tube is straight).
In this way the rolling angle of a short segment can be detected.
Unfortunately, the periodicity in the equatorial line, which provides
the diameter, can not be measured. Even if the indices can not be
determined, CBED may be useful for an analysis of the rolling angles
of carbon nanotubes. The angle can be measured from the relative distances
as indicated in Fig. \ref{cap:CBED}.

\subsection{Diffraction on related nanotube structures}

Diffraction is possible on a wide range of nanotube structures. Using
the same experimental conditions, we have obtained diffraction patterns
also from double-walled nanotubes (DWNTs), multi-walled nanotubes
and small bundles (not shown). In double-walled nanotubes, multi-walled
nanotubes, and nanotube bundles, multiple orientations of the graphene
lattice are visible in a diffraction pattern. Considerations for assigning
the indices of DWNTs can be found in \cite{KociakDWNTaccuracy03}.
An index assignment for bundles and MWNTs is often not unique, as
it is not clear which tube diameter belongs to a specific orientation.

\section{Conclusions and outlook}

A diffraction analysis of carbon nanotubes is presented, including
a novel sample preparation method, description of simulation methods,
and electron diffraction at an energy below the threshold for knock-on
damage. Experimental parameters are described and discussed, permitting
the reader to perform a similar analysis (e.g. statistics of tube
indices). We have shown single-nanotube diffraction in the Köhler
illumination condition as well as convergent-beam electron diffraction.
The goal of this study is a reliable determination of the structural
indices for a suspended nanotube in our free-standing structures.
This is achieved for the majority of the candidate nanotubes.

The importance of a reliable, non-destructive analysis of the nanotube
indices lies in combining the structural analysis with other experiments
on the same object. This becomes possible through our lithographically
defined free-standing structure. We are currently collecting data
from single-tube transport measurements and Raman spectroscopy using
similar sample structures as shown here \cite{MeyerKB05proc,MeyerEDRam05}.
These are, finally, measurements on well-defined molecules that allow
a correlation of electronic and vibrational properties with structural
information. We point out that the free-standing structure can be
designed almost arbitrarily by lithography so that complex free-standing
structures, including nanotubes, can be prepared for a variety of
novel experiments.

\begin{acknowledgments}
The authors acknowledge financial support by the BMBF project INKONAMI
(Contract No. 13N8 402) and the EU project CANAPE. We thank xlith.com
for lithography services. We thank Prof. M. Ruehle and his group for
support with TEM, especially C. Koch for very helpful discussions.
M. P. acknowledges support from Dr. B. Chaudret's group, Dr. V. Jourdain,
Dr. P. Poncharal and Prof A. Zahab.
\end{acknowledgments}
\bibliographystyle{unsrt}
\bibliography{bibtex/books,bibtex/TEM,bibtex/diverse,bibtex/QM,bibtex/Raman}

\end{document}